\documentclass[11pt,twoside]{article}
\usepackage{./asp2014}
\newcommand{\zav}[1]{\left(#1\right)}
\newcommand{\hzav}[1]{\left[#1\right]}

\newcommand{\hvezda}{BS\,Cir}
\aspSuppressVolSlug
\resetcounters

\DeclareMathAlphabet{\mathsc}{OT1}{cmr}{m}{sc}

\DeclareRobustCommand{\ion}[2]{%
\relax\ifmmode
{\mathbf{#1\,\mathsc{#2}}}
\else\textup{#1\,{\sc{#2}}}%
\fi}

\begin{document}

\title{Preliminary study of the moderately cool chemically peculiar star BS Circini}
\author{Zden\v{e}k Mikul\'a\v{s}ek,$^{1}$ Jan Jan\'ik,$^{1}$ Ji\v{r}\'{i} Krti\v{c}ka,$^{1}$ Miloslav Zejda,$^{1}$  and Miroslav~Jagelka$^{1}$}
\affil{$^1$Department of Theoretical Physics and Astrophysics, Masaryk University, Brno, Czech Republic; \email{mikulas@physics.muni.cz}}

\paperauthor{Sample~Author1}{Author1Email@email.edu}{ORCID_Or_Blank}{Author1 Institution}{Author1 Department}{City}{State/Province}{Postal Code}{Country}
\paperauthor{Sample~Author2}{Author2Email@email.edu}{ORCID_Or_Blank}{Author2 Institution}{Author2 Department}{City}{State/Province}{Postal Code}{Country}
\paperauthor{Sample~Author3}{Author3Email@email.edu}{ORCID_Or_Blank}{Author3 Institution}{Author3 Department}{City}{State/Province}{Postal Code}{Country}

\begin{abstract}
BS Cir is a representative of moderately cool magnetic chemically peculiar stars which displays very strong light variations in Str\"omgren index $c_1$ indicating large changes in the height of the Balmer jump. We present two-spot model of light variations fitting successfully all of nine light curves obtained in the spectral region 335-750~nm. We also discuss the nature of the observed variations of intensities of Fe, Cr, Ti, Si, Mg and RE spectral lines and possible mechanisms matching the observed light variations. It was confirmed that the observed period of BS Cir $2.204$ d is rising with the rate of $\dot{P}=5.4(4)\times 10^{-9}$. The found minor secular changes in the shape of light curve should be compatible with the period changes caused by precessional motion due to magnetic distortion of the star.
\end{abstract}

\section{Introduction}
Variable flux redistribution from far ultraviolet to long-ward spectral regions qualitatively explains the light variability of the majority of mCP stars with extended abundance spots. Such photometric spots (effectively one or two) are generally bright in near ultraviolet, optical, and infrared regions \citep[see e.g.][]{krt901,krtEE,krtcu,shulyak}. However, some moderately cool mCP stars of SrCrRE type share also spots being dark in $v$ and its spectral vicinity. Symptomatic for these CP stars are large amplitudes in the Str\"omgren index $c_1=(u-v)-(v-b)$.

Representatives of such type of mCP stars are  `Maitzen's star' - CS~Virginis = HD\,125\,248, `Mikulasek's star' - CQ~UMa = HD\,119\,213, 'Olsen's star' - BR~Crucis = HD\,110\,956, the star with the maximum amplitude among all of mCPs, VV~Scl = $\alpha$~Scl\,=\,HD\,7676, and \hvezda\ = HD\,125\,630, which is the southern analogue of CS\,Vir \citep[for details and light curves see in the `Zoo of mCP light curves in \textit{uvbyHp}',][]{zoo}.

These stars are not intensively studied at present and the nature of their extravagant variability remains unknown up to now.
\begin{figure}[h]
\centerline{\includegraphics[width=0.70\textwidth]{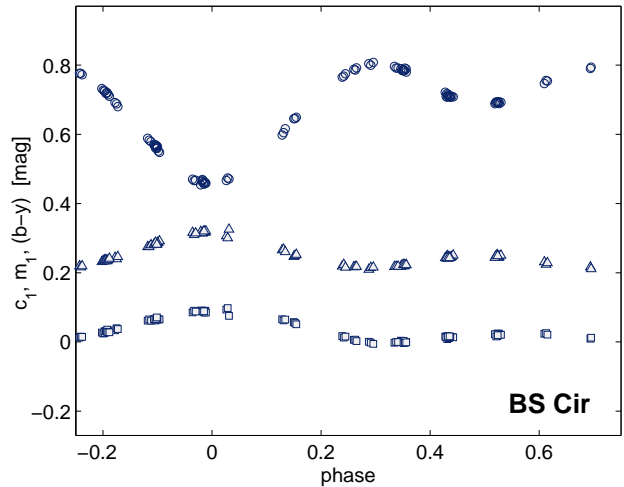}}
\begin{center}
\caption{\small The phase variations of Str\"omgren colour indices $c_1,\ m_1, \mathrm{and}\ (b-y)$ of the moderately cool magnetic chemically peculiar star BS Circini show very strong variability of the first index expressing a height of the Balmer jump. Changes of next two indices are in antiphase with $c_1$ variations. The nature of those variations remains mysterious up to now. The phase graph was constructed by means of \textit{uvby} observations of \citet{mare83}} \label{c1index}
\end{center}
\end{figure}

\section{The basic characteristics of \hvezda}

\hvezda\,=\,HD\,125630\,=\,HIP\,70346, is a moderately cool magnetic chemically peculiar star of the SrCrEu type with the modest rotational period $P = 2\fd20423$ \citep[][]{mik13,mik15}. \hvezda\ has strong antiphase light variations in $y$ and $v$ (see Fig.\,\ref{krivky}), large $c_1$ amplitude (see Fig.\,\ref{c1index}).

Combining all available kinematic, photometric and spectroscopic data about the star, we derived the following astrophysical parameters:
$T_{\mathrm{eff}}=8800\pm500\,\mathrm{K},\
L=41.7\pm1.4\,\mathrm{L}_{\odot},\ M=2.32\pm0.14\,\mathrm{M}_{\odot},\ \mathrm{and}\
\mathrm{age}= 510^{+90}_{-150}$\,Myr. The star has a moderate global magnetic field of bipolar strength $B_{\rm{p}}$ of several kG \citep{kobacu,hub06}.

\section{The photometry and the period of \hvezda}

\hvezda\ was observed in 1975-6 by \citet{vofa79} in $\textit{uvby}$ and then in 1980 by \citet{mare80,mare83} who revealed the star to be photometrically variable with a period $P=2\fd205\pm0\fd004$ \citep[apparently not taking photometry of] [in their account]{vofa79}. Data set of \citet{mare83} was later reanalysed and the found period was confirmed by \citet{mama85}.

\citet{cale93} then added \citet{mare83} data to their 56 precise $\mathit{uvby}$ measurements taken in 1991 and determined the period to $P'=2\fd20552(6)$. As they did not consider \citet{vofa79} data they could be mistaken in the total number of cycles between 1980 and 1991 ($\Delta t \simeq 11\times 365\simeq4015$\,d) by one
$(\Delta k=-1)$. This false period is given in later catalogues and articles excluding the period list of \citet{dubath} containing automatically found periods and types of variable stars in the Hipparcos survey. \hvezda\ is there announced as an eclipsing binary with the period of 1\fd1020 (the half of the true value).

\begin{figure}[h]
\centerline{\includegraphics[width=0.75\textwidth]{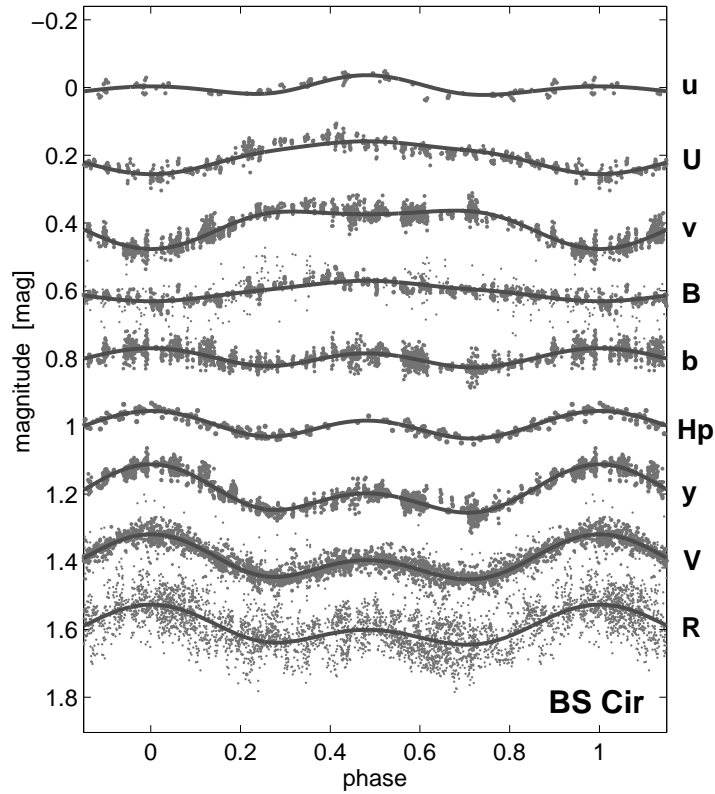}}
\begin{center}
\caption{\small Light curves of \hvezda\ in various filters arranged according their effective wavelengths. The shapes of light curve can be well interpreted by the model with two photometric spots centered at phases 0.0 and 0.48 (see Eq.\,\ref{model}) with the contrasts depending on the effective wavelengths. The rotational phase is calculated according to the ephemeris
with a quadratic term (Eq.\,\ref{efemerida}). The area of individual full circles are equal to their weights. The sources of all used photometric data  plot are listed in Tab.\,\ref{tab1}.} \label{krivky}
\end{center}
\end{figure}

\begin{figure}[h]
\centerline{\includegraphics[width=0.98\textwidth]{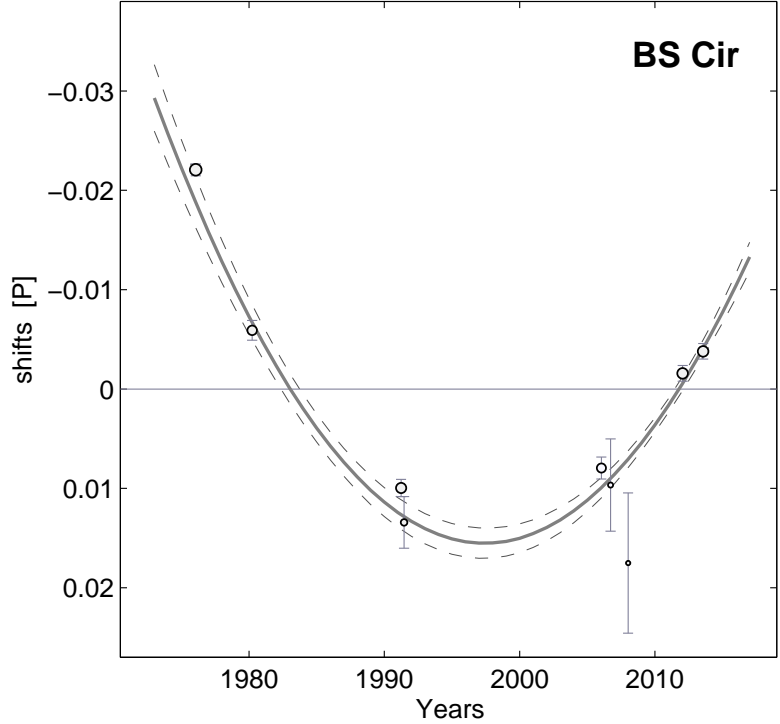}}
\begin{center}
\caption{\small The phase shifts of observed light curves versus light curves predicted by the linear ephemeris ($\vartheta-\vartheta_0$), see Eq.\,\ref{efemerida}, Tab.\,\ref{tab1},  can be well fitted by a parabola. It indicates that the observed period of light variations is lengthening more or less linearly with the rate: $\dot{P}=5.4\times10^{-9}$. The mean period $P_0=2\fd2042849(4)$. The one-$\sigma$ deflections from the fit is denoted by dashed lines, the areas of individual markers are inversely proportional to their uncertainty signed also by error bars.} \label{OC}
\end{center}
\end{figure}

The true period should be given by the relation: $P=P'+\Delta k\,P'^2/\Delta t\doteq2\fd2043$. This period was then fully confirmed by our period analysis of all above mentioned data sets replenished by Hipparcos \citep{esa}, ASAS \citep{pojm}, Integral \citep{omc}, and Pi of the Sky \citep{burd} survey photometries. Nevertheless residuals calculated versus to the linear fit O-C$_{\rm{lin}}$ signed that the period may change (see, Sec.\,\ref{kvadrat}, Fig.\,\ref{OC}). It was one of the reasons why we observed \hvezda\ at South African Astronomical Observatory by various photometric instruments in 2011-14. Here we obtained 7124 individual observations in \textit{bvy} and \textit{UBV} photometric systems that confirmed our suspicions.

\begin{figure}[h]
\centerline{\includegraphics[width=0.86\textwidth]{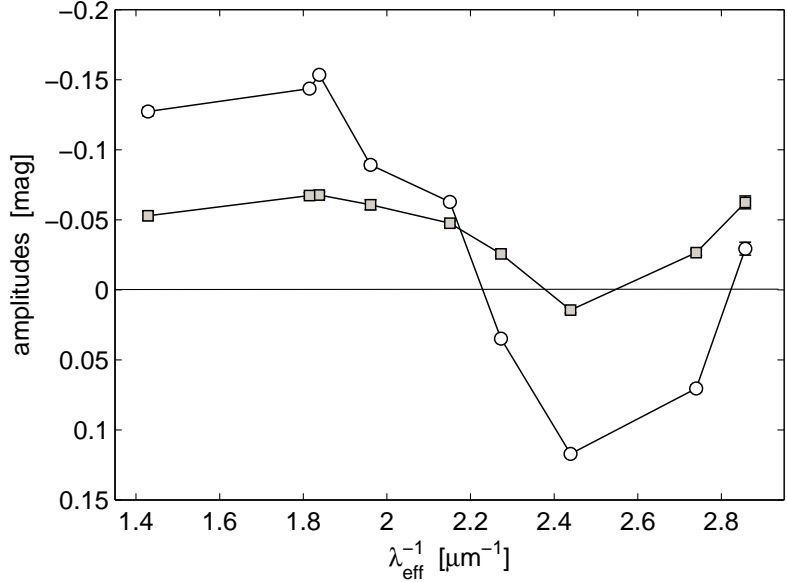}}
\begin{center}
\caption{\small The dependence of amplitude of the primary (circles) and secondary (squares) photometric spots on the effective wavelengths of individual colours. It shows that the primary spot is dark in the wavelength interval 355 -- 445 nm, while the secondary one is dark only in the close vicinity of the center of the $v$ color (410~nm). The spots are bright in other parts of spectrum covered by photometric observations.} \label{Fig3}
\end{center}\label{ampl}
\end{figure}

Presently, \hvezda\ is one of the best photometrically monitored mCP stars. We have in our disposal observations more or less evenly covering the interval 1975-2014 (39 years), the total number of individual observations, taken in 9 filters 350-700 nm, is 13\,591, the average uncertainty of them is $\sigma = 0.018$ mag. Light curves in individual colours, plotted according to quadratic phase (see Sec.\,\ref{kvadrat}), display very  disparate shape (see Fig.\,\ref{krivky}). Nevertheless, we have ascertained that all of them can be well described by a simple  phenomenological model assuming the presence of two uneven symmetric photometric spots centered to phases $\varphi_{01} = 0.0000(5)$ and $\varphi_{02} = 0.4803(11)$, with half-widths $d_1=0.14,\ d_2=0.12$:
\begin{eqnarray}
&\displaystyle m(\lambda,q)= m_{0q}+\sum_{k=1}^2\, A_k(\lambda)\,\hzav{\exp\zav{1-\cosh \frac{\varphi_k}{d_k}}-2.2809\,d_k},\label{model}\\
&\varphi_k=(\varphi-\varphi_{0k})-\mathrm{round}(\varphi-\varphi_{0k}),\nonumber
\end{eqnarray}
where $m(\lambda,q)$ is the model prediction of the magnitude of observation done in the filter of the effective wavelength $\lambda$, $A_1(\lambda)\ \mathrm{and}\ A_2(\lambda)$ are the amplitudes of both photometric spots in a photometric colour with the wavelength $\lambda$, $m_{0q}$ is the mean magnitude of the $q$-th set of photometric observations, $d_1,d_2$ are the half-widths of the light curves of relevant photometric spots expressed in phases, $\varphi$ is the phase determined according to quadratic ephemeris (see Eq.\,\ref{efemerida}), and $\varphi_{01}, \varphi_{02}$ are the phases of the centers of the spots.

The model allowed to use all available photometric data and served as the general model for the fine period diagnostics described in Sec.\,\ref{kvadrat}. Dependence of amplitudes on the effective wavelengths indicates that light curves of the star in individual colours are the result of at least two different variability mechanisms.

\section{Spectrum. Mechanisms of light variability}

Our research of the spectrum of \hvezda\ is based on the analysis of 30 spectrograms obtained by us in 2011-12 at La Silla Observatory using spectrograph FEROS $(R=48\,000)$ attached to the 2.2m MPG/ESO telescope.  The optical spectrum of the moderately cool mCP star $(T_{\mathrm{eff}} < 9000$\,K) \hvezda\ is rich in lines of \ion{H}{i}, metals \ion{Fe}{i,\,ii}; \ion{Cr}{i,\,ii}; \ion{Ti}{i,\,ii}; \ion{Si}{i,\,ii}; \ion{Mg}{ii}; and rare earths. All lines are heavily blended, spectrum is completely covered by lines and no continuum is visible. Line spectrum is apparently variable in intensity and line profile.

The brief inspection of spectrograms covering substantial part of a phase diagram led us to the finding that phase curves of line intensities of the most of chemical elements mimic light curves (double waves, dominant the primary spot), however relative changes in intensities of metals are relatively small and very likely insufficient to produce the observed pronounced changes in light. The agreement between phase curves of metal line intensities is not perfect - we see here a systematic shift of 0.1 of the period. The strongest variations exhibit \ion{Eu}{ii} lines,  phase curves of which can be expressed by the LC model (see Eq.\,\ref{model}) without any shift.

We speculate that there are two causes of the observed \hvezda\ light variations: 1)~redistribution of the stellar flux from far UV to near UV and the optical region due to uneven abundances of metals, 2) changes in the magnitude of the Balmer jump due to variable pollution of the atmosphere by metals and namely RE \citep[see][]{zoo,shulyak}.

\section{\hvezda\ -- next mCP star with an unstable period}\label{kvadrat}

The prevailing majority of mCP stars display constant periods and stable shapes of light curves what implies that these stars are rotating as a solid body and surface structures stabilized by strong surface global magnetic field is persistent in time scales of centuries. Nevertheless, there were revealed a few of altogether hot mCP stars as very hot $\sigma$ Ori\,E \citep{town} and V901 Ori \citep{mik07,mik901,mik11,mik13}, or hot fastly rotating CU Vir \citep{pyper04,pyper13,mik11,mik13} exhibiting almost continuous variations in their observed periods \citep{unstead}. We \citep{mik11,mik13} speculate in the last two mentioned stars that dynamic interactions between a thin, outer magnetically-confined envelope, braked by the stellar wind, with an inner, faster rotating stellar body are able to explain the observed rotational variability. The lengthening of the $\sigma$ Ori\,E rotational period can be caused by the magnetic braking.

The deep period analysis of \hvezda\ photometric data was surprising: we revealed well documented linear growth of the observed period of light changes with the rate of $\dot{P} = 5.4(4) \times 10^{-9}$. It means that during the last 39 years the period of the star lengthened by $6.7(5) \mathrm{s}/39$ yr. The second period derivative ($\ddot{P}$ is nearly zero). We have derived the quadratic orthogonal ephemeris for the phase function $\vartheta(t)$ which origin was put in the weighted center of our photometric data:
\begin{eqnarray}\label{efemerida}
& \displaystyle \vartheta_0(t)=\frac{t-M_0}{P_0};\quad \vartheta(\vartheta_0)=\vartheta_0-\textstyle{\frac{1}{2}}\,\dot{P}
\zav{\vartheta_0^2-a_1\,\vartheta_0-a_2}; \\
& \displaystyle E=\mathrm{floor}(\vartheta);\quad \varphi=\vartheta-E;
\quad a_1=\frac{\overline{\vartheta_0^3}} {\overline{\vartheta_0^2}}=-2.99;
\quad a_2=\overline{\vartheta_0^2}=3.24,
\end{eqnarray}
where $\vartheta_0,\ \mathit{\Theta},\ a_1,\ a_2$ are auxiliary variables or constants, $M_0$ is the selected origin of the phase function $\vartheta(t)$ for linear approximation: $M_0=2\,453\,943.59154$, the mean period $P_0=2\fd2042849(4)$, $\dot{P} = 5.4(4) \times 10^{-9}$. $E$ is the common epoch and $\varphi $ is the quadratic phase.

\begin{table}
\begin{center}
\caption{The list of sources of individual photometric data of \hvezda\ contains: the mean HJD datum of measurements, O-C in days versus the quadratic ephemeris (\ref{efemerida}), the used filters, the number of observations, and the citation of the source. }
\begin{tabular}{|c l c r l|}
  \hline
mean JD&\ \ \ O-C [d]&filters&number&source\\
\hline
2\,442\,800&-0.003\,4(6)&\textit{uvby}&24&\citet{vofa79}\\
2\,444\,323&\ 0.000\,8(9)&\textit{uvby}&391&\citet{mare83}\\
2\,448\,343&-0.002\,6(8)&\textit{uvby}&52&\citet{cale93}\\
2\,448\,423&\ 0.000\,7(23)&\textit{BHpV}&744&\citet{esa}\\
2\,453\,749&-0.001\,9(11)&\textit{V}&1088&\citet{pojm}\\
2\,453\,998&\ 0.000(4)&\textit{V}&732&\citet{omc}\\
2\,454\,472&\ 0.010(7)&\textit{R}&3436&\citet{burd}\\
2\,455\,945&-0.001\,2(7)&\textit{vby, UBV}&4804&\citet{mik13}\\
2\,456\,499&\ 0.003\,4(7)&\textit{vby, UBV}&2320&this paper\\
\hline
\end{tabular}\label{tab1}
\end{center}
\end{table}

The LiTE period changes due to variable radial velocity caused by orbital motion in the system of 2 bodies (spectroscopically invisible companion) can be excluded because we have not noted any variations in radial velocity of the star.

\citet{shade} predicted precession of magnetically distorted stars. It should cause also marginal cyclic of light curves and period variations. \citet{mik901,mik11,mik13} showed that the constancy of the light curves shapes and the extent of the observed  period variations of V901 Ori and CU Vir evidently overcoming the limit allowed by the precessional model \citep[see the Appendix of the article of][]{mik901} excluded the precession as the explanation for period changes of these notorious mCP stars. Nevertheless, total variation of \hvezda\ in phase is not so large: $0.11\,P$, consequently we cannot excluded that it might be caused by the stellar precession. This explanation is supported by finding of small long-term changes in shape of light curve (amplitude of both spots decreases) while the phase distance between centers of the spots remains constant.

\section{Conclusions}
\begin{itemize}
\item We observed the moderately cool southern mCP star \hvezda\  both photometrically and spectroscopically.
\item We improved the ephemeris of light variations and successfully modeled light curves in the region 350-750 nm by two nearly opposite different photometric spots.
\item We revealed changes in the depths of \ion{Cr}{ii}, \ion{Fe}{ii}, \ion{Mg}{ii}, and \ion{Eu}{ii} lines correlated with light variations. The nature of photometric spots of moderately cool mCP~stars was briefly discussed.
\item Using all available photometric data we have found well documented lengthening of the observed period and long-term changes in \hvezda\ light curve shapes. The facts could be explained by the free precession of the aspheric star deformed by its magnetic field.
\end{itemize}

\acknowledgements This paper uses observations made at the South African Astronomical Observatory (SAAO). The research is also based on observations obtained at the European Southern Observatory (ESO programmes 087.D-0099 and 089.D-0153). This work was funded by the grant GA\v{C}R P209/12/0217.

\end{document}